\documentclass[a4paper,twocolumn]{esapub2005} 
\usepackage{times}
\usepackage{natbib}
\usepackage{graphicx}
\usepackage{psfig}

\title{INTEGRAL/IBIS 20-100 keV Extragalactic survey: an update}
\author{L. Bassani, A. Malizia and J.B. Stephen}
\affil{On behalf of the INTEGRAL AGNs survey team}
\affil{IASF-Bologna/INAF, via P. Gobetti 101, 40129 Bologna, Italy}

\begin{document}


\maketitle

\begin{abstract}
  
  Analysis of INTEGRAL Core Program and public Open Time observations
  has recently provided a sample of 60 extragalactic sources
  selected in the 20-100 keV band above a flux of 1.5 10$^{-11}$ erg
  cm$^{-2}$ s$^{-1}$. As this band probes heavily obscured
  regions/objects, i.e. those that could be missed in optical, UV, and
  even X-ray surveys, our sample offers the opportunity to study
  the extragalactic sky from a different point of view with respect to
  surveys at lower energies.  We present an update of our analysis,
  including the first sample of AGNs detected above 100 keV.  We also
  discuss the results of follow up observations performed at optical
  and X-ray frequencies with the aim of classifying our objects and
  studying the effects of intrinsic absorption in gamma-ray selected
  AGNs. The average redshift of our sample is 0.134 while the mean
  20-100 keV luminosity in Log is 43.84; if blazars are excluded these
  numbers become 0.022 and 43.48 respectively.  Defining an absorbed
  object as one with N$_{H}$ above 10$^{22}$ atoms cm$^{-2}$, we find
  that absorption is present in 60\% of the objects with at most 14\% of the
  total sample due to Compton thick active galaxies.  Almost all
  Seyfert 2s in our sample are absorbed as are 24\% of Seyfert 1s.  We
  also present broad-band spectral information on a sub-sample of the
  brightest objects: our observations indicate a  mean photon
  index of $\Gamma$=1.8 spanning from
  30-50 keV to greater than 200 keV. Finally, we discuss the LogN/LogS
  distribution in the 20-100 and 100-150 keV bands derived from our
  sample.  The present data highlight the capability of INTEGRAL to
  probe the extragalactic gamma-ray sky, to discover new AGNs and to
  find absorbed objects.
\end{abstract}

\section{Introduction}

Quantifying the fraction of AGNs missed by surveys which are affected
by selection due to absorption is necessary if we want to fully
understand the accretion history of the Universe and study a
population of objects (absorbed AGNs) so far poorly explored.
Furthermore, the measurement of the primary AGN continuum and its
cut-off energy is crucial for understanding emission models and
discriminating between them.  Both information (spectral shape of AGN
and column density distribution) are key parameters for estimating the
contribution of AGNs to the X-ray cosmic diffuse background and for
testing current unified theories. Only high energy observations (above
10 keV) can provide an unbiased knowledge of the column densities and 
high energy cut-off distributions of AGNs in the local Universe.\\
A step forward in both these issues is now provided by INTEGRAL/IBIS 
and SWIFT/BAT which are surveying a great fraction of the sky above 20
keV with a sensitivity better than a few mCrab and a point source
location accuracy of the order of 1-3 arcminutes depending on the
source strength. These two surveys are complementary to each other, as
the first one concentrates mostly on mapping the galactic plane,
while the second covers the high galactic latitude sky 
so that together they will provide the best yet sample of AGNs 
selected in the gamma-ray band.\\
We have recently presented a catalogue of $\sim$ 60 IBIS AGNs selected
in the 20-100 keV band above a flux limit of 
1.5$\times$10$^{-11}$ erg cm$^{-2}$ s$^{-1}$ (\citep{ba06} or
paper I).  The analysis was performed using INTEGRAL Core Program and
public Open Time observation performed up to April 2005.  Most sources
in the sample were Seyfert galaxies almost equally divided between
type 1 and 2 objects, 6 were blazars but 14 were still unclassified at
the time of this first publication.  Despite an initial attempt to
study the role of absorption in the sample, a significant fraction
(about 30\%) of the objects did not have archival X-ray spectra to
allow a full assessment of the column density distribution. 
Furthermore due to the highly inhomogeneous coverage of the INTEGRAL/IBIS 
survey a determination of the LogN-LogS relation had to await a proper study 
of the various corrections to be applied.\\
Here, we present an update on the first 20-100 keV IBIS catalogue of
AGNs, with a number of identification/classification being provided in
the meantime through optical spectroscopy; on the basis of further
work we have eliminated a few sources from the initial list and added
a couple of new objects.  We also use available X-ray data to obtain
information on the column density of a number of AGNs in the
catalogue.  Given the higher number of new information available, we
have re-estimated the main conclusions of \citep{ba06}.
Furthermore on the basis of this update, we have evaluated the
LogN-LogS relation in the 20-100 keV band.  In the meantime, we have
also performed broad band spectral analysis of a subsample of objects
observed in the first INTEGRAL extragalactic survey \citep{ba04}
using IBIS data in combination with archival X-ray data from
other missions and have provided information on their primary
continuum shape including the cut-off energy.
Finally, we report the first catalogue of AGNs detected above 100 keV
by INTEGRAL and provide the first extragalactic number counts in the
100-150 keV band.

\section{Catalogue update and optical classification}

Of all the excesses reported in paper I, 3 objects resulted a
posteriori to be artifacts of the imaging processes mainly due to
projection effects; these 3 sources (IGRJ 13000+2529, IGR J13057+2036 and
UGC3142) have been removed from the initial sample. Two objects which
were not listed but were detected in the survey, were found to be
AGNs a posteriori through optical follow up observations (\citep{ma06}
and references therein); due to the limited
information available on these two excesses (IGR J17488-3253 and IGR
J17513-2011) their AGNs nature was difficult to assess prior to
optical spectroscopic work. It is important to stress that some other
unidentified sources may turn out to be active galaxies but due to
their location near the galactic plane it is difficult to assess their
nature without a proper optical follow up work.  We have estimated for
these two objects the total exposure and 20-100 kev flux and
luminosity as done in paper I for the other sources; ISGRI coordinates
can instead be found in \citep{bi06}. We obtained an exposure
of 737.0 and 571.8 ks, a flux of 2.18$\pm$0.17 and 2.04$\pm$0.22 mCrab
and a luminosity of 3 and 15.5 $\times$ 10$^{43}$ erg cm$^{-2}$ s$^{-1}$ for
IGR J17488-3253 and IGR J17513-2011 respectively.  We remind that
between 20-100 keV, 1 mCrab corresponds, for a Crab like spectrum, to
1.6 $\times$ 10$^{-11}$ erg cm$^{-2}$s$^{-1}$ which is a value close to
our detection limit; the fluxes have been converted to gamma-ray
luminosities assuming H$_0$=71 Km sec$^{-1}$ Mpc$^{-1}$ and q$_0$=0
\citep{sp03}.  Finally one source (IGR J16194 -2810) has been
removed as it is not clear at the present stage from our own follow up
measurements (Masetti, private communication) if it is a galactic or an
extragalactic source.  A number of objects in the initial catalogue
have been optically classified and confirmed to be active galaxies; at
the moment only 5 objects have no classification nor redshift
available.  For all the objects in this initial survey, we report in
Table 1 an update of the source classification and redshift either
available from the literature or from our own work (\citep{ma06}
at this conference and references therein); in some cases the optical 
classification has been revised allow intermediate Seyfert types to be also sampled.  
We have also updated
information on the column density by using available X-ray data.  In
particular we have made use of public SWIFT/XRT observations
and performed our own spectral analysis in order to estimate photon
index and column density for a set of 8 sources. More detailed results
of this analysis are presented in the contribution to this conference
by Landi et al. \citep{la06}. We have also used XMM data obtained very recently in
conjunction with INTEGRAL data to quickly estimate the source column
density. A more in depth analysis of these objects is on going but
since we are only interested in estimating if the column density is
above or below 10$^{22}$ at cm$^{-2}$ a simple modeling of the data
(power law absorbed both by galactic and where required by intrinsic absorption)
is sufficient to provide this information. Only in the
case of IC4518A  we find that an intrinsic column density above 10$^{22}$
at cm$^{-2}$ is required by the data.  Results becoming avaliable in
the literature have also been used to update our database \citep{pi06}.
Only 9 objects have no column density estimate at this
stage.

\begin{table}
  \begin{center}
    \caption{AGNs   sample}\vspace{1em}
    \renewcommand{\arraystretch}{1.2}
    \begin{tabular}[h]{lrcc}
      \hline
      Source & Type      &  z          & Log Nh \\
      \hline
QSOB0241+62           & S1     & 0.044 & 22.2        \\
MCG+8-11-11           & S1.5   & 0.020 & $<$20.3      \\
IGR J07597-3842       & S1.2     & 0.040 & $<$20.6$^{\dagger}$     \\
ESO 209-12            & S1.5   & 0.040 & $<$21.4 $\ddagger$           \\
Fairall 1146          & S1.5   & 0.031 & 21.1 $\ddagger$              \\
MCG-05-23-016         & S2     & 0.008 & 22.2         \\
IGR J10404-4625       & S2     & 0.024 & $>$22.0      \\
NGC4151               & S1.5   & 0.003 & 22.47         \\
4C04.42               & Bl     & 0.965 & -             \\
NGC4388               & S2     & 0.008 & 23.63         \\
3C273                 & Bl     & 0.158 &  $<$20.5       \\
NGC4507               & S2     & 0.012 & 23.5          \\
LEDA 170194           & S2     & 0.037 & 22.28        \\
NGC4593               & S1     & 0.009 & $<$20.3    \\
3C279                 & Bl     & 0.536 & $<$20.3     \\
NGC4945               & S2     & 0.002 & 24.60         \\
CenA                  & S2     & 0.002 & 23.4      \\
4U1344-60             & S1.5   & 0.013 & 22.6$^{\star}$    \\
IC4329A               & S1.2   & 0.016 & 21.6    \\
Circinus              & S2     & 0.001 & 24.6    \\
IGR J16482-3036       & S1     & 0.031 & 21.1$^{\dagger}$\\
ESO138-G01$^b$        & S2     & 0.009 & $>$24.0    \\
NGC6300               & S2     & 0.004 & 23.5    \\
GRS1734-292           & S1     & 0.021 & $>$21.7   \\
2E1739.1-1210         & S1     & 0.037 & 21.2$\ddagger$\\
IGR J17488-3253       & S1     & 0.02  & 21.3$^{\dagger}$    \\
IGR J17513-2011       & S1.9     & 0.046 & --    \\
IGRJ18027-1455        & S1     & 0.035 & 21.4$\ddagger$   \\
PKS1830-211$^c$       & Bl     & 2.507 & 21.4   \\
ESO103-G35            & S2     & 0.013 & 23.2   \\
2E1853.7+1534         & S1.2   & 0.084 & --  \\  
NGC6814               & S1.5   & 0.005 & $<$20.7   \\
Cygnus A              & S2     & 0.056 & 23.6    \\
IGRJ21247+5058        & S1?    & 0.020?&  --  \\
1ES0033+595           & Bl     & 0.086 & 21.6    \\
NGC788                & S2     & 0.014 & 23.3   \\
NGC1068               & S2     & 0.004 & $>$25.0  \\
NGC1275$d$            & S2     & 0.018 & 22.2    \\
3C111                  & S1    & 0.049 & $<$21.9   \\
LEDA168563             & S1    & 0.029 & -   \\
MKN3                   & S2    & 0.014 & 24.0    \\
MKN6                   & S1.5  & 0.019 & 23.0    \\
IGR J07565-4139        & S2   & 0.022?& 22.0    \\
QSO0836+710            & Bl    & 2.172 & $<$20.5   \\
IGR J12026-5349        & S2    & 0.027 & 22.3  \\
IGR J12415-5750        & S2    & 0.024 & 21.0$^{\dagger}$   \\
ESO323-G077            & S1.2  & 0.015 & 23.7    \\
MCG-06-30-15           & S1.2  & 0.008 & $<$20.3    \\
ESO511-G030            & S1    & 0.022 & $<$20.7   \\
IGR J14492-5535        & -     & -     & 23.0$^{\dagger}$    \\
IGR J14552-5133        & NLS1    & 0.018 & -    \\
      \hline \\
      \end{tabular}
  \end{center}
\end{table}

\begin{table}
  \begin{center}
   \centering{Tab 1 - Continued} 
   \renewcommand{\arraystretch}{1.2}
    \begin{tabular}[h]{lrcc}
      \hline
      Source & Type      &  z          & Log Nh \\
      \hline

IGR J16119-6036        & S1    & 0.016 & -    \\
IC4518-A$^o$           & S2    & 0.016 & 23.3$\ddagger$    \\
NGC6221$^b$            & S2    & 0.005 & 22.0    \\
IGR J16558-5203        & S1.2    & 0.054 & $<$20.0    \\
IGR J17204-3554        & -     & -     & 23.1\\
IGR J18249-3243        & -     & -     & -    \\
IGR J20187+4041        & -     & -     & 23.3    \\
IGR J20286+2544        & S2    & 0.014 & 23.6-24.0    \\
IGR J21178+5139        & -     & -     & -   \\

      \hline \\
      \end{tabular}
    \label{tab:table}
  \end{center}
$\dagger$ Landi et al. these proceedings\\
$\ddagger$ Panessa et al. in preparation and De Rosa et al. in preparation\\
$\star$ Piconcelli et al. (2006), also absorption partially covering the source\\
\end{table}

\begin{figure}
\includegraphics[width=8.0cm,height=8.0cm]{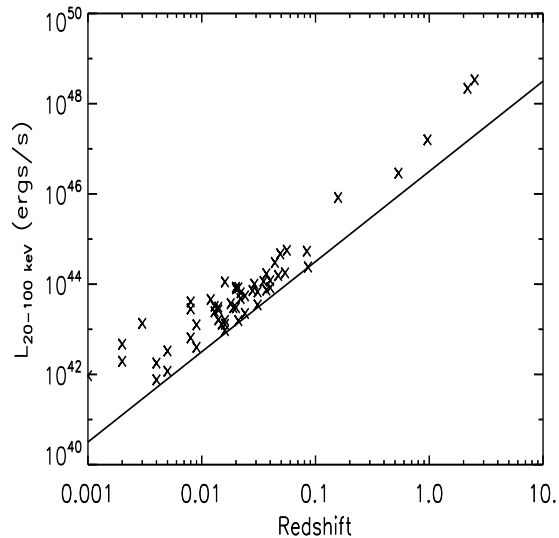}
\caption{20-100 keV Luminosity versus redshift for optically classified AGNs    in the present survey; 
straight line corresponds to the IBIS survey limit}
\end{figure}

\section{Results update}

Our extragalactic survey comprises after this revision 60 objects
securely identified with AGNs.  For those objects with known distance,
we plot in figure 1 the gamma-ray luminosity against redshift, to show
the large range in these parameters sampled by the present survey.
>From this figure it is also evident that our sensitivity limit is
around 1.5 $\times$ 10$^{-11}$ erg cm$^{-2}$s$^{-1}$ (straight line in
the figure).
Within the overall sample, 49 objects are classified as Seyfert
galaxies, 6 are blazars and only 5 (8$\%$ of the sample) are still
unclassified. Within the sample of Seyfert galaxies, 26 objects are of
type 1-1.5 while 23 are of type 2, i.e. a ratio 1:1, which illustrates
the power of gamma-ray surveys to find narrow line AGN.  About 10$\%$
of our sample is made of radio loud objects.
The average redshift of our sample is 0.134 while the mean 20-100 keV luminosity
in Log is 43.84; if blazars are excluded, these numbers become 0.022
and 43.48 respectively.\\
The column density distribution for our sources is presented in figure
2.  Assuming 10$^{22}$ atoms cm$^{-2}$ to be the dividing line between
absorbed and unabsorbed sources, we find that absorption is present in
nearly $60\%$ of the sample of objects having  a column density estimate. 
This result is in line with the Swift findings \citep{mk05} 
and also with other INTEGRAL AGNs   surveys \citep{be06}, \citep{sa06}.

\begin{figure}[th!]
\includegraphics[width=8.0cm,height=8.0cm]{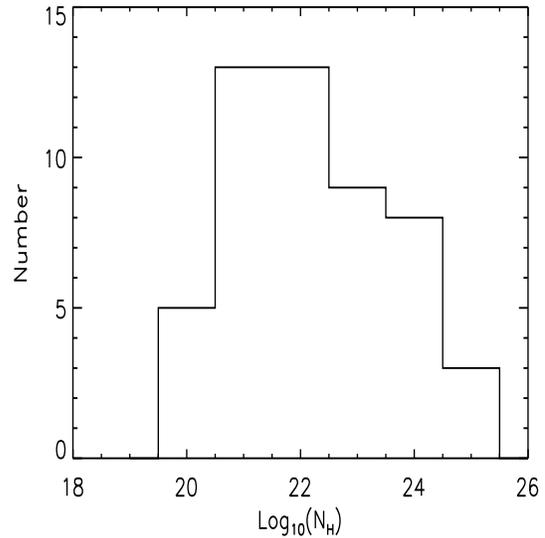}
\caption{Column density versus 20-100 keV Luminosity for AGNs in the survey with known intrinsic absorption }
\end{figure}

\begin{figure*}[th!]
\begin{center}
\vspace{-1.cm}
\mbox{\psfig{file=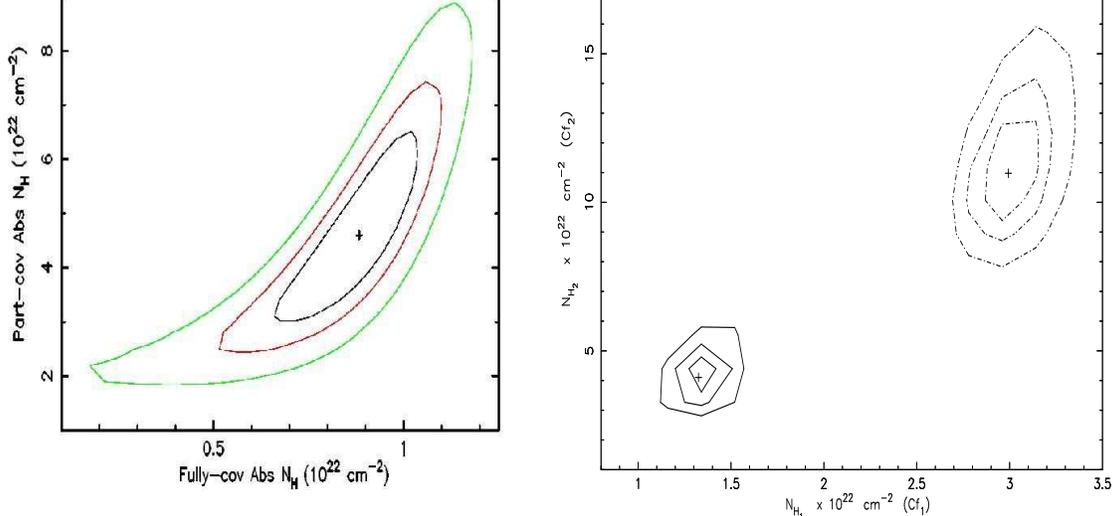,width=7.0cm,height=7cm}}
\hspace{1.0cm}
\mbox{\psfig{file=fig3b.ps,width=7.0cm,height=7cm,angle=-90}}
\caption{Contours of column densities in 4U 1344-60 (one fully versus one partially covering
the source) and in MKN 6 (in this case both partially covering the source) ASCA and BeppoSAX data clearly 
indicating N$_{H}$ variability.}
\end{center}
\end{figure*} 

Within the present sample we find that 6 possibly 7 objects are Compton
thick. NGC4945, Circinus galaxy, NGC1068 and MKN3 are well known
examples of this class \citep{pi06}. To this set we add ESO 138-G01 {\footnote {analysis of various 
energy band mosaics indicates detection of both ESO138-G01 and 
NGC 6221, which are 11' apart}} \citep{cb00}, possibly NGC6300 since this is an
object which moves from thick to thin in time \citep{mg03} and very likely
IGR J20286+2544 \citep{la06}, \citep{ma06}. Therefore we estimate that at most 14$\%$ of the sample is
Compton thick.  Within the sub-sample of 22 Seyfert 2s with known
N$_{H}$, we find that 95 $\%$ are absorbed and  30\% are Compton
thick; this is in line with previous estimates of the column density
distribution of type 2 objects based on X-ray data \citep{ri99}, \citep{ba99}.
Interestingly we also find that 24\%
of type 1-1.5 objects with N$_{H}$ values are absorbed.
In some of these absorbed type 1 objects, the absorption is complex with multiple column density
obscuring partially and/or totally the central nucleus; in same cases
these absorptions could also vary in time.  
Typical examples of this complexity are MKN 6, NGC4151 and
4U1344-60 (see \citep{pi06} and references therein).
In figure 3 we show the contours of the column densities for 4U1344-60
\citep{pi06} and MKN 6 \citep{ma03}. 
Similar sources are also found in the SWIFT/BAT survey \citep{lr06}.\\
In figure 4 we show the column density of our sources as a function of
IBIS luminosity. Although from this figure there is little evidence of
any strong correlation, the few high luminosity objects in the survey
all have low absorption. However, since high luminosity objects are
mostly blazars, this evidence is not conclusive and more data are
required to firmly establish this observational evidence.\\
More in depth studies of the present sample require optical
classification of all objects and a detailed analysis of their broad
band behavior (particularly in the X-ray to gamma-ray band) in order
to fully understand the role of absorption : in particular we want to
establish if there is any other Compton thick object in the sample and
how is the absorption (complex or not) in Seyfert 1-1.5 galaxies.
Furthermore when all sources are properly characterized it will be
possible to estimate the luminosity function and the role of gamma-ray
selected AGNs with respect to the cosmic diffuse background (see also \citep{sa06}).

\begin{figure}[th!]
\includegraphics[width=8.0cm,height=8.0cm]{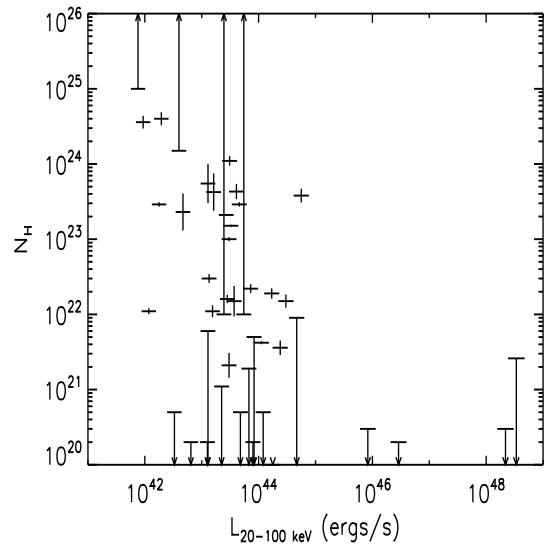}
\caption{Column density versus 20-100 keV luminosity for AGNs in the present sample}
\end{figure}

\begin{figure*}[th!]
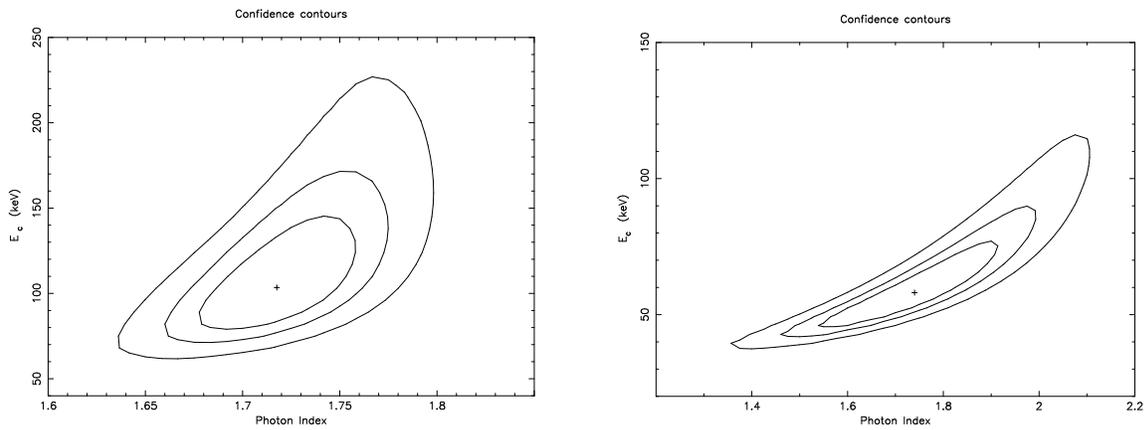

\begin{center}
\mbox{\psfig{file=fig5a.ps,width=7cm,angle=-90}}
\hspace{1.0cm}
\mbox{\psfig{file=fig5b.ps,width=7cm,angle=-90}}
\caption{High Energy cut-off vs photon index for MCG-5-23-16 (right) and GRS 1734-292 (left).}
\end{center}
\end{figure*}

\section{Study of the primary continuum}

The high energy emission of AGNs is often to
first order well described by a power law of photon index 1.8-2.0,
extending from a few keV to over 100 keV; at higher energies there is
evidence of an exponential cut-off.  Secondary features, which are
also commonly present, are considered to be the effects of
reprocessing of this primary continuum and are relatively well
understood.  Modeling of high energy AGN spectra has so far
generally focused on how to reproduce and explain the observed
primary continuum shape: however, while the photon index distribution has 
been well investigated \citep{ma01}, observational results on the cut-off energy have so far
been limited by the scarcity of measurements above 10-20 keV, with
most information coming from BeppoSAX broadband spectra. Analysis of
type 1 and 2 AGN \citep{pe02} provides evidence for a wide range of
values in the cut-off energy, spanning from 30 to 300 keV and further
suggests a possible trend of increasing cut-off energy as the power
law index increases; it is not clear however if this effect is due to
limitations in the spectral analysis or if it is intrinsic to the
source population sampled.\\

\begin{figure}[th!]
\centering \includegraphics[width=8.0cm,height=8.0cm]{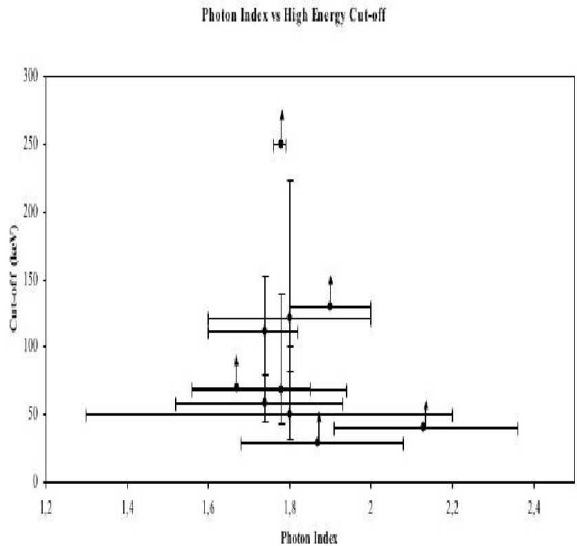}
\caption{Power law photon index versus  cut-off energy for the sample of AGNs analyzed 
by Soldi et al. 2005 and Molina et al 2006.}
\label{cutoff}
\end{figure}

Direct INTEGRAL measurements enable us to obtain further observational
results on the primary continuum and high energy cut-off thus
providing more refined parameters for AGN modeling. The sample of 12
AGNs first detected by INTEGRAL \citep{ba04} can be taken as a case
study: they are representative of the larger sample of AGNs   presented
here and are detected with adequate statistics to allow more in depth
analysis. In particular, they are ideal objects for the study of the
primary power law component, the presence of any high energy cut-off
and the existence of a relation (if any) between these two parameters.
Spectral analysis has recently been reported for this entire sub-sample using
INTEGRAL data alone or in conjunction with data from other satellites
\citep{so05}, \citep{de05}, \citep{mo06}.
Excluding PKS1830-211 which is a blazar, this sub-sample is primarily made of Seyfert galaxies.
Indeed PKS 1830-211 shows a flat spectrum and a cut-off likely located
at MeV energies as typical of a low frequency peaked or red blazar \citep{de05}.
For the remaining sources, the mean photon index
is 1.83$\pm$0.07, while the cut-off energy ranges from 30-50 keV to
greater than 200 keV. In most cases the cut-off energy is well
constrained at or below 100 keV: the contour plot of photon index
versus cut-off energy for MCG-05-23-16 and GRS 1734-292
are shown in figure 5 and indicates the ability by high energy observations 
of putting tight constraints on both parameters .  
Combining all our data, we can also test the correlation found by a number of authors 
between the photon index and the cut-off energy.  In figure 6, we plot the photon index versus
the high energy cut-off for those Seyfert galaxies in the sub-sample for
which we have information on these two parameters.  It is clear from
this figure that our data do not show the correlation found by
\citep{pe02}: the values for the photon index, in fact, cluster around
1.8, while the values of the cut-off energy range over a wide
interval. We can conclude that the correlation between the high energy
cut-off and the photon index remains to be proved and that a diversity
in cut-off energy is most likely a property of Seyfert galaxies.
INTEGRAL will keep observing these AGNs   and new ones  for the duration of
the mission and so we expect that more spectral data on this class of objects
will become available; this will provide further information on the
high energy cut-off in AGNs and a deeper insight into the primary
continuum emission.

\section{AGNs detected above 100 keV}

Recently Bazzano et al. \citep{bz06}  have presented the first census of IBIS
objects with detection above 100 keV.  This high energy survey
includes 10 AGNs with detection above 4 sigma level in the 100-150 keV
band of which only one (Cen A) is also visible in the 150-300 keV band.
All objects are also detected at energies below 100 keV.  Of these 10
AGNs, 2 are blazars and 8 are Seyfert galaxies; however this time the
ratio of type 1 versus type 2 objects is 2:6, with Seyfert 2
outnumbering Seyfert 1 by a factor of 3, which is interestingly close
to values found in optical spectroscopic surveys \citep{ma95}, \citep{ho97}.
However, this could be simply due to the
limited number of AGNs present in this first extragalactic survey above
100 keV.  In Figure 7, we plot for these 10 AGNs the hardness ratios H1 versus H2,
where H1 is defined as the 40-100 keV to 20-40 keV flux and HR2 as the
ratio of the 100-150 keV to the 40-100 keV flux. It is clear from this
figure that the 10 AGNs   detected at high energies have photon index
$\ge$ 1.5 with some clearly showing a softening of the spectrum probably
due to the presence of a cut-off at or around 100 keV as discussed in
the previous section.

\begin{figure}
  \centering \includegraphics[width=8cm]{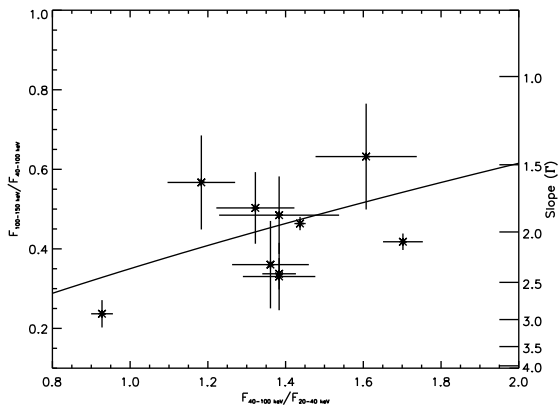}
\caption{HR1 (40-100 keV/20-40 keV) versus HR2 (100-150 keV/40-100 keV) for the 10 AGNs detected above 100 keV}
\label{cutoff}
\end{figure}

\section{AGN  LogN-logS in the 20-100 keV and 100-150 keV bands}

In this section, we estimate the number counts relation for AGNs   in two
bands : 20-100 keV and 100-150 keV. Even though the number of sources
in the second band is low, a LogN-LogS estimate is nevertheless
a useful information to have if one wants to estimate the total number
of objects which may become visible as the survey progresses.  However
before doing any evaluations, there are a number of complications
involved in forming this relationship that must be taken into
consideration. First, the sensitivity limit in any direction depends
not only on the exposure at that point, which is extremely non-uniform due to the
pointing strategy of INTEGRAL \citep{bi06}, but also on systematic variations/structures in 
the background.  Another problem is
the fact that our survey is not truly serendipitous as some of our
objects have been the targets of INTEGRAL dedicated observations.
Finally, our sample is probably incomplete again due to the peculiar
nature of INTEGRAL observing strategy.  Given these caveats, we first
convert exposure to 1 $\sigma$ limiting sensitivity the flux errors
and effective exposures for the sources found in each band. 
(see fig 8 for the case where the errors refer to 20-100 keV fluxes).
From this we obtain the sky coverage as a function of 5$\sigma$ (4$\sigma$) 
20-100 keV (100-150 keV) limiting flux
as shown in figure 9, again for the case of 20-100 keV band.  

\begin{figure}
  \includegraphics[width=8.0cm,height=8.0cm]{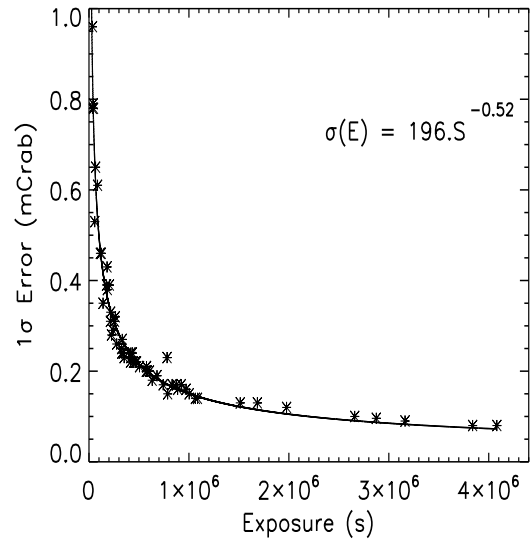}
\caption{The 1$\sigma$ background variation as a function of exposure in the 20-100 keV band.}
\end{figure}

\begin{figure}
  \includegraphics[width=8.0cm,height=8cm]{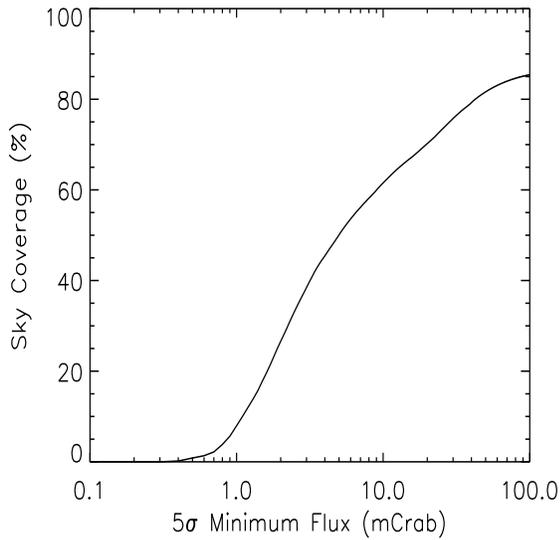}
\caption{The percentage of sky coverage as a function of 5$\sigma$ limiting flux.}
\end{figure}

We then create the `raw' logN-logS for the
60 AGNs detected in the 20-100 keV band (figure 10) and the 10 found in the
100-150 keV band (figure 11).  This distribution is then corrected for the
exposure limit and sky coverage assuming {\em a-priori} an isotropic underlying
distribution.  Both the `raw' data and the corrected values
normalized to full sky coverage are also shown in figures 10 and 11 where the
horizontal bars indicate the errors on the flux of the N$^{th}$
source.
The best-fit power law relationship and the (1$\sigma$)
statistical limits are indicated by the solid lines.
 Unsurprisingly, we find that the best-fit power law is similar to, though 
rather flatter than an isotropic distribution. A deeper analysis will be performed as a
greater number of sources and, most importantly, more uniform sky coverage become available.

Finally these number counts can be used to estimate the contribution of local
AGNs to the cosmic diffuse background.
The total AGNs   emissivity measured in both bands above 1 mCrab accounts for approximately 1-3$\%$ 
of the total background intensity; extrapolation of our LogN-LogS by a
factor of 100 towards lower flux will account for 15-20$\%$ of the
extragalactic background. In short we are just start to unveil the
source population responsible for the hardX/gamma-ray background.

\begin{figure}
  \includegraphics[width=8.0cm,height=8cm]{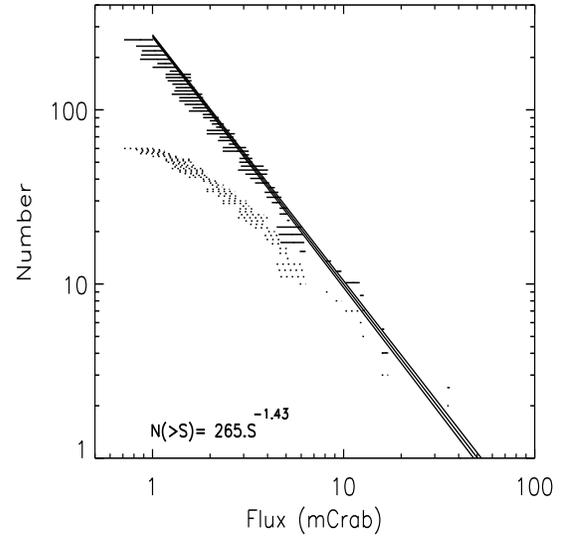}
\caption{20-100 keV full sky number-flux relationship for our sample}
\end{figure}

\begin{figure}
  \includegraphics[width=8.0cm,height=8cm]{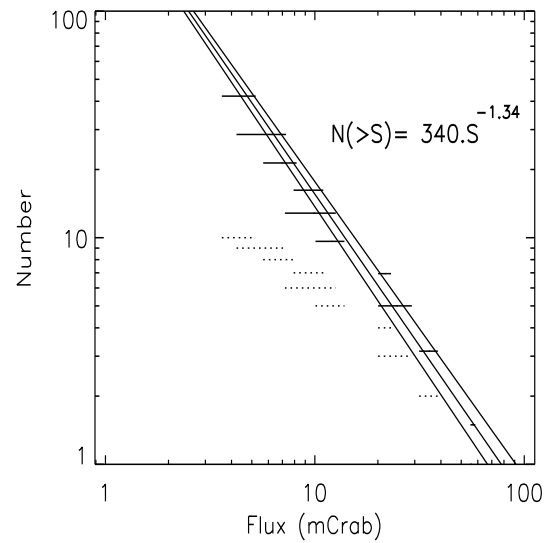}
\caption{100-150 keV full sky number-flux relationship for the sample of 10 AGNs reported in \citep{bz06}.}
\end{figure}

\section{Future expectations}

Future work will concentrate on the optical classification
of objects in the present sample and of those  new AGNs that will be detected in
the future. We also want to progress in estimating the column
densities for the 8 remaining sources in the 20-100 keV catalogue and prepare to obtain X-ray data
on the new sources. In parallel, we expect to concentrate on detailed
analysis of the spectral properties of all AGNs in the present sample
in order to define the primary continuum (photon index and cut-off
energy) using INTEGRAL data in combination with archival X-ray
information and data obtained though proposal requests to the Chandra, XMM
and Swift observatories.  In figure 12 we plot the number of AGNSs
detected in our own previous INTEGRAL surveys \citep{ba04}, \citep{bi06},
\citep{ba06} as a function of science windows
analyzed.  We are currently working at the third INTEGRAL survey which
involves the analysis of around 24000 science window.  As evident in
the figure we expect around 100 AGNs (even more as new part of the
extragalactic sky will be observed for the first time) to be detected
in this coming survey thus doubling the number of sources for which follow up
optical and X-ray observations will be needed, investigation of their
spectral properties required and statistical studies performed. The
future looks good for extragalactic studies with INTEGRAL and busy for
the observers.

\begin{figure}
  \includegraphics[width=8.0cm,height=8cm]{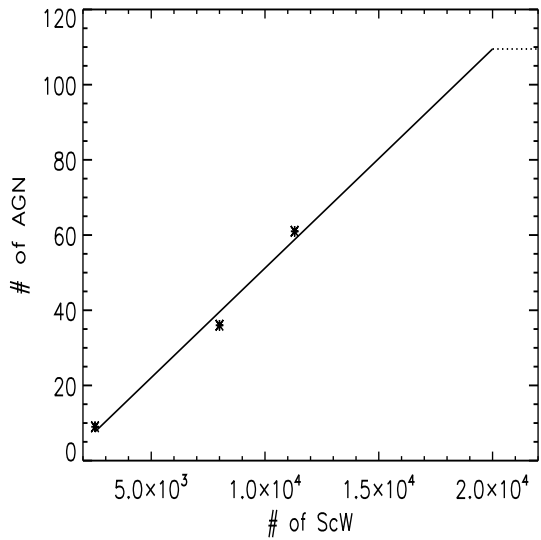}
\caption{The increase in AGN numbers as a function of accumulated science windows. When the 3$^{rd}$ IBIS/ISGRI survey becomes available around 110 AGN should be visible.}
\end{figure}

\section{Acknowledgements}
This research has been supported by ASI under contract I/R/046/04 and
I/023/05. The authors thank all their collaborators for their continuum efforts and their 
entusiasm.
This research has made use of data obtained from NED (Jet
Propulsion Laboratory, California Institute of Technology), SIMBAD
(CDS, Strasbourg, France) and HEASARC (NASA's Goddard Space Flight
Center).

\end{document}